\documentclass[conference]{IEEEtran}
\IEEEoverridecommandlockouts

\usepackage{cite}
\usepackage{amsmath,amssymb,amsfonts}
\usepackage{graphicx}
\usepackage{textcomp}
\usepackage{xcolor}
\usepackage{url}
\usepackage[T1]{fontenc}

\def\BibTeX{{\rm B\kern-.05em{\sc i\kern-.025em b}\kern-.08em
T\kern-.1667em\lower.7ex\hbox{E}\kern-.125emX}}

\begin{document}

\title{Asymmetric Decline Rather Than Backfire: \\A Repost-Network Analysis of Japanese Vaccine-Shedding Misinformation on X}

\author{
\IEEEauthorblockN{Akinobu Seki}
\IEEEauthorblockA{
\textit{The University of Tokyo}\\
Tokyo, Japan\\
seki@torilab.net}
\and
\IEEEauthorblockN{Erina Murata}
\IEEEauthorblockA{
\textit{The University of Tokyo}\\
Tokyo, Japan\\
murata@torilab.net}
\and
\IEEEauthorblockN{Masaki Chujyo}
\IEEEauthorblockA{
\textit{The University of Tokyo}\\
Tokyo, Japan\\
mchujyo@g.ecc.u-tokyo.ac.jp}
\and
\IEEEauthorblockN{Fujio Toriumi}
\IEEEauthorblockA{
\textit{The University of Tokyo}\\
Tokyo, Japan\\
tori@sys.t.u-tokyo.ac.jp}
}

\maketitle

\begin{abstract}
This study investigates the diffusion of vaccine-shedding discourse on X in 2024, focusing on polarization in repost networks and reactions to corrective information. Here, ``shedding'' refers to the scientifically unsupported claim that harmful substances are released from individuals vaccinated against COVID-19 and cause adverse health effects in nearby unvaccinated people. Using 109,170 posts containing the Japanese term for ``shedding'' and 478,089 reposts collected from January 1 to December 31, 2024, we constructed a post-level repost network, identified discourse clusters, and examined changes in repost participation before and after major events. The results show that shedding discourse formed a two-pole structure consisting of pro-vaccine and anti-vaccine clusters. After news coverage denying shedding, we found no evidence of diffusion-level backfire in anti-vaccine repost activity. Instead, although the overall topic declined, repost participation in the anti-vaccine cluster decreased more sharply than that in the pro-vaccine cluster. User transition analysis further suggests that this change was not driven by the conversion of strongly anti-vaccine users, but by the entry of previously non-expressive users into pro-vaccine diffusion and the withdrawal of anti-vaccine supporters from repost activity. These findings suggest that corrective information may reshape the composition of misinformation diffusion participants without necessarily converting core opponents.
\end{abstract}


\section{Introduction}

Vaccination is one of the most important public health measures for preventing severe illness and death from infectious diseases \cite{larson2014,macdonald2015}. In recent years, social media has become a major channel for vaccine-related information, creating an environment in which scientifically unsupported claims and conspiracy theories can spread rapidly and widely \cite{cinelli2020,wilson2020}. 
The spread of vaccine-related misinformation on social media is particularly consequential because vaccine hesitancy has been identified as a major global health challenge, and misinformation is considered one of its key drivers \cite{larson2014,macdonald2015}.
In the broader context of online misinformation, large-scale studies have shown that false news can diffuse more rapidly and broadly than true news \cite{vosoughi2018}. 
Vaccine-specific studies further demonstrate the public health relevance of vaccine-related misinformation.
Exposure to online misinformation about COVID-19 vaccines has been shown to reduce vaccination intent, and regional-level misinformation prevalence has been associated with lower vaccination rates \cite{loomba2021,pierri2022}.

Although corrective information from experts, news organizations, and government agencies has been positioned as an important countermeasure, effective correction remains challenging.
Nyhan et al. showed in a randomized controlled trial that corrective messages can reduce vaccine-related misperceptions without necessarily increasing vaccination intent \cite{nyhan2014}. Chan and Albarrac\'{\i}n's meta-analysis further clarified that the success of correction depends on topic and context \cite{chan2023}, while psychological mechanisms that make misinformation resistant to correction have also been discussed \cite{ecker2022}. How corrective information diffuses through social media networks and whether it may be followed by diffusion-level backfire therefore remain important questions. In this study, we use the term \textit{diffusion-level backfire} to refer to renewed or increased anti-vaccine repost activity following corrective information. Unlike the traditional backfire effect, which refers to situations in which corrective information reinforces individual-level misperceptions or anti-vaccine beliefs, this concept concerns changes in repost-based diffusion behavior.

The ``shedding'' discourse on Japanese X in 2024 provides an empirical case for examining these questions. In this context, shedding refers to the claim that harmful substances are released from individuals vaccinated against COVID-19 and cause health damage to unvaccinated people nearby. The term vaccine shedding originally refers to a technical concept discussed in relation to vaccines such as live attenuated vaccines, in which pathogen-derived components may replicate in the body; it does not directly apply to standard mRNA vaccines. The Japanese Ministry of Health, Labour and Welfare also states that, regarding replicon vaccines, there is no scientific evidence that vaccine components are transmitted from vaccinated individuals to others \cite{mhlw_qa2024}. 

Despite this lack of scientific support, shedding discourse became salient in Japan in October 2024, around the start of the routine COVID-19 vaccination program on October 1 and subsequent public responses to shedding claims. This timing provides a useful observational setting because the discourse was accompanied by both public attention and corrective information, including the joint statement by three academic societies---the Japanese Association for Infectious Diseases, the Japanese Respiratory Society, and the Japanese Society for Vaccinology---denying shedding, and related news coverage on October 21 \cite{jacso2024,nhk2024}.

In this study, we examine shedding discourse on Japanese X in 2024 using a repost-network approach. Specifically, we identify the community structure of the repost network surrounding shedding discourse, analyze changes in diffusion behavior around the start of routine vaccination and subsequent corrective responses, and examine transitions in user stance classifications based on repost behavior. By focusing on a specific misinformation theme in Japanese-language social media, this study aims to clarify how misinformation and corrective information recirculated across communities and whether corrective information was followed by diffusion-level backfire.

This study makes three contributions. First, it provides an issue-specific analysis of Japanese shedding discourse and identifies a clear two-pole repost-network structure. Second, it examines diffusion-level backfire and finds an asymmetric decline in repost activity following corrective information. Third, by combining network clustering with user stance transition analysis, it shows that the post-event change was driven mainly by the withdrawal of anti-vaccine users and the entry of previously non-expressive users, rather than by the conversion of core opponents.


\section{Related Work}

\subsection{Vaccine Discourse on Social Media}

Research on vaccine discourse on social media has developed along two main lines: content analysis and network-structure analysis. Johnson et al. showed that pro-vaccine, anti-vaccine, and undecided groups compete in online spaces, and that anti-vaccine groups have structural advantages in reaching undecided users \cite{johnson2020}. Pierri et al. showed that vaccine misinformation on Twitter was driven in part by low-credibility sources and a small number of superspreader accounts \cite{pierri2023}. Toriumi et al. showed that, in Japanese COVID-19 vaccine discourse, exposure to anti-vaccine discourse was associated with changes in political representation \cite{toriumi2024}.

\subsection{Network Analysis and Community Detection}

In vaccine research, network analysis based on repost relations has become an important method for detecting communities. To identify such community structures, the Louvain method \cite{blondel2008} is widely used for this purpose. Rao et al. showed that retweet graphs reflect echo chambers more strongly than follow graphs, because reposts involve an active choice of whom to amplify \cite{rao2023}. At the same time, Guerra et al. warned against interpreting retweets as agreement, noting that they are also used for critical or antagonistic resharing \cite{guerra2017}. Repost networks are thus useful not only for detecting homogeneous communities but also for observing diffusion-level responses to corrective information, including possible increases in repost activity among opposing groups.

\subsection{Misinformation Correction and Backfire Effects}

In misinformation-correction research, the possibility that corrections may backfire has long been an important concern. Nyhan et al. showed in a randomized controlled trial that corrective messages can reduce vaccine-related misperceptions without necessarily increasing vaccination intent, highlighting the limitations of correction \cite{nyhan2014}. However, subsequent evidence has led to more cautious interpretations of the backfire effect. Chan and Albarrac\'{\i}n's meta-analysis did not support the view that corrections are generally counterproductive, instead showing that their success depends on the topic, context, and recipient characteristics \cite{chan2023}.

Meanwhile, whether corrective information reaches broad audiences on social media is a separate issue. Corrections may remain confined within homogeneous networks or be quoted and shared antagonistically by opposing groups. Such dynamics have also been discussed in studies on the resistance of misinformation beliefs to correction \cite{ecker2022}. Therefore, whether correction leads to diffusion-level backfire should be tested empirically in relation to network structure.

\subsection{User Stance Transitions}

Although many studies have examined changes in stance proportions at the aggregate level, relatively few have shown that the same users frequently shift from strongly anti-vaccine to pro-vaccine positions. Vaccine attitudes have relatively stable psychological foundations, and conspiracy thinking, anti-elitism, concerns about bodily purity, and individualistic worldviews have been shown to be associated with anti-vaccine attitudes across 24 countries \cite{hornsey2018}. Nevertheless, longitudinal research indicates that vaccine acceptance can vary with perceived infection threat, policy developments, and social influence \cite{fridman2021}, and such changes are more likely to occur among intermediate, peripheral, or newly participating users than among strong core supporters.

Taken together, these studies provide the basis for analyzing vaccine discourse, repost-network structure, correction, and stance dynamics. The present study builds on these strands of research by examining how misinformation and corrective information recirculated across communities in Japanese-language shedding discourse.

\section{Data and Methods}

\subsection{Data Collection}
This study analyzes posts on X that contained the Japanese term for ``shedding'' between January 1 and December 31, 2024, together with reposts of those posts. The data were collected using the X API. The dataset includes original post IDs, timestamps, post text, user IDs, usernames, repost counts, repost IDs, repost user IDs, and repost usernames. Timestamps were standardized to Japan Standard Time and aggregated by day and by analysis period.

This study distinguishes between original posts and reposts of those posts. Posters are defined as unique users who created original posts, and repost users are defined as unique users who reposted them. Because the repost data do not include the time at which each repost was made, period-based repost aggregations assign each repost to the period corresponding to the timestamp of the original post.

\subsection{Event Detection}
\label{sec:Event Detection}

To understand the temporal dynamics of shedding discourse, we calculated daily post counts. We also computed a 7-day moving average to reduce weekday effects. We then analyzed annual trends in post volume and identified periods of major increases in posting as peaks.

Next, to examine whether these peaks corresponded to specific social events, we calculated frequent terms in the posts around each peak. By comparing these frequent terms with actual news reports and official materials, we examined the correspondence between post-volume peaks and social events. Through this procedure, we identified periods in which shedding discourse changed substantially and used them as reference points for before-and-after analyses.

\subsection{Repost Network Construction and Clustering}

Prior studies have shown that repost behavior in vaccine-related discourse tends to be polarized by stance \cite{toriumi2024}. Based on this finding, we assumed that shedding discourse would also form diffusion communities based on repost behavior. Following the methods of Uchida et al. \cite{uchida2019} and Toriumi et al. \cite{toriumi2024}, we constructed a post-level network based on repost behavior and performed clustering.

To construct the repost network, we included posts that had been reposted at least twice and represented each post as a node. For each post $i$, let $U_i$ denote the set of users who reposted post $i$. The similarity between posts $i$ and $j$ was calculated using the Simpson coefficient:
\begin{equation}
S_{ij} = \frac{|U_i \cap U_j|}{\min\left(|U_i|, |U_j|\right)}.
\label{eq:simpson}
\end{equation}
This value represents the extent to which the repost-user sets overlap, normalized by the smaller of the two sets. In this study, an edge was placed between posts when $S_{ij} \geq 0.3$, and an undirected, unweighted network was constructed.

We applied the Louvain method \cite{blondel2008} to the resulting post-level network to extract post clusters. The Louvain method detects communities by maximizing modularity $Q$ \cite{newman2004}, which is defined as
\begin{equation}
Q = \frac{1}{2m}\sum_{i,j}\left(A_{ij}-\frac{k_i k_j}{2m}\right)\delta(c_i,c_j),
\label{eq:modularity}
\end{equation}
where $A_{ij}$ is the adjacency matrix, $k_i$ and $k_j$ are the degrees of nodes $i$ and $j$, $m$ is the number of edges, $c_i$ is the community assignment of node $i$, and $\delta(c_i,c_j)$ is 1 if nodes $i$ and $j$ belong to the same community and 0 otherwise. Larger values of $Q$ indicate that edges are more concentrated within communities than expected under a degree-preserving random network.

After extracting clusters, we evaluated similarities between clusters based on the overlap in their repost-user sets. Cluster overlap was calculated as the number of shared repost users divided by the smaller of the two cluster-level repost-user sets. Cluster pairs were merged in descending order of this overlap rate, and the merging process stopped when no pair had an overlap rate exceeding 50\%.

Finally, for each merged cluster, we calculated frequent terms in the included post texts and interpreted the discourse characteristics. When the assumed two-pole structure was obtained, the cluster denying shedding was labeled the pro-vaccine cluster, and the cluster endorsing or promoting shedding was labeled the anti-vaccine cluster.

\subsection{Event-Based Comparison of Repost Diffusion}
\label{sec:Event-Based Comparison of Repost Diffusion}

We evaluated changes in repost diffusion around the key events by comparing the number and composition of repost users in the pro-vaccine and anti-vaccine clusters. For each period, we counted unique users who reposted posts originally published in that period, separately for each cluster. These counts are non-exclusive: users who reposted posts from both clusters originally published in the same period were counted in both cluster totals.

Equal-length analysis periods were defined based on the event dates identified in Section~\ref{sec:Event Detection}. We compared periods immediately before and after each event to capture short-term changes, and also compared pre-event periods with later post-event periods to distinguish immediate reactions from subsequent changes.

Changes in the number of repost users in each cluster were evaluated
using negative binomial regression, a standard approach for overdispersed
count data \cite{hilbe2011,agresti2013}. Daily repost-user counts on
social media are overdispersed, with counts spiking sharply on viral days.
Negative binomial regression accommodates this structure via a dispersion
parameter $\alpha$: specifically, if $Y$ denotes the daily unique
repost-user count and $\mu = \mathrm{E}[Y]$ denotes its expected value,
the variance is given by $\mathrm{Var}(Y) = \mu + \alpha\mu^2$, where
$\alpha$ is the dispersion parameter estimated from the data. This model
has been widely adopted in prior social media studies
\cite{hilbe2011,sutton2015,chuai2026}.

We modeled daily unique repost-user counts as the outcome variable with 
a binary period indicator (comparison period = 1, reference period = 0) 
as the sole predictor, fitted by maximum likelihood using 
\texttt{statsmodels} in Python \cite{seabold2010}. The exponentiated coefficient yields the rate ratio (RR): RR $> 1$ indicates an increase and RR $< 1$ a decrease. For chi-square tests of the anti-vaccine user ratio, we used mutually exclusive user categories: users who reposted only anti-vaccine cluster posts were counted as anti-vaccine users, users who reposted only pro-vaccine cluster posts were counted as pro-vaccine users, and users who reposted posts from both clusters originally published in the same period were excluded from the contingency tables. This ensured that each remaining user contributed to only one cell within each period. All tests used a significance level of 5\%.

\subsection{User Stance Classification and Transition Analysis}

To examine which user groups contributed to the changes in repost-user counts and anti-vaccine repost-user ratios described in Section~\ref{sec:Event-Based Comparison of Repost Diffusion}, we classified users based on their repost behavior and analyzed their transitions across periods. This classification does not directly measure users' attitudes, but captures which side of the discourse each user redistributed through reposting.

For each period, we counted each user's reposts of pro-vaccine and anti-vaccine cluster posts and calculated the corresponding pro-vaccine and anti-vaccine repost ratios. Users whose pro-vaccine repost ratio was 90\% or higher and whose anti-vaccine repost ratio was 10\% or lower were defined as pro-vaccine supporters. Among them, users with three or more total reposts in the period were defined as strong pro-vaccine supporters, while those with one or two reposts were defined as weak pro-vaccine supporters. Similarly, users whose anti-vaccine repost ratio was 90\% or higher and whose pro-vaccine repost ratio was 10\% or lower were defined as anti-vaccine supporters. Among them, users with three or more total reposts were defined as strong anti-vaccine supporters, while those with one or two reposts were defined as weak anti-vaccine supporters.

Users whose pro-vaccine and anti-vaccine repost ratios both fell between 10\% and 90\% were defined as intermediate users. Users who belonged to the user set for a given transition analysis but did not repost in a given period were defined as non-expressive users. This classification was applied to each period, and user stance transition matrices were constructed between periods. This allowed us to examine which groups moved after the events and whether transition patterns differed between users with strong and weak repost-based stances.

\section{Results}

\begin{figure}[t]
  \centering
  \includegraphics[width=\columnwidth]{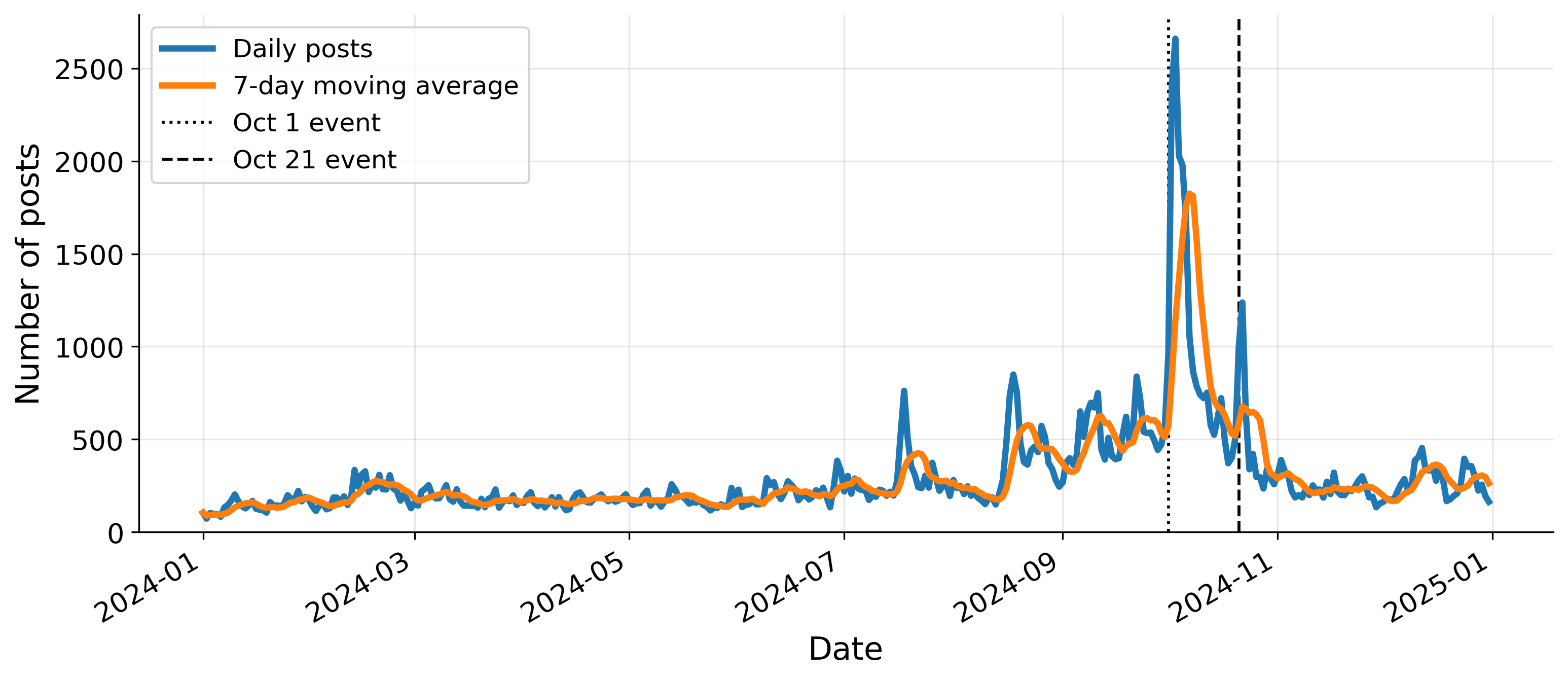}
  \caption{Daily post counts and 7-day moving average of shedding-related posts in 2024. The dotted and dashed vertical lines indicate the two event dates used in the analysis: October 1, corresponding to the start of Japan's routine COVID-19 vaccination program, and October 21, corresponding to the academic-society statement and related media coverage denying shedding. The figure shows a sharp increase in post volume after October 1 and a smaller, subsequent increase around October 21.}
  \label{fig:daily}
\end{figure}

During the target period, we collected 109,170 posts containing the Japanese term for ``shedding,'' posted by 24,671 unique users; these posts received 478,089 reposts from 87,288 unique repost users. Building on this dataset, we first identify the event dates from the temporal trend of posts, then construct and interpret the repost network, and finally analyze event-based changes in
repost diffusion and user stance transitions.

\subsection{Event Detection Results}

Figure~\ref{fig:daily} shows the daily post counts and the 7-day moving average. Post volume fluctuated throughout the year, with notable increases in early October and around October 21.

Although October 1 was not the day with the highest post count, it corresponded to the start of the 2024 routine COVID-19 vaccination program. After October 1, the number of posts increased rapidly, reaching 2,411 on October 2 and the highest daily count of 2,660 on October 3. The 7-day moving average peaked at 1,825 on October 7. The total number of posts during the preceding 7 days (September 24--30) was 3,580, whereas the total during the following 7 days (October 1--7) was 12,775, an approximately 3.6-fold increase. Comparing the frequent terms during this period with news reports and official materials suggested that the increase corresponded to the start of routine COVID-19 vaccination, including the replicon vaccine (Table~\ref{tab:event_terms}). Therefore, this study treats October 1 as the event date marking the beginning of the post increase associated with the start of routine vaccination \cite{mhlw2024}.

After the large early-October surge subsided, post volume increased again, with 1,000 posts on October 21 and 1,237 on October 22. The total number of posts during the preceding 7 days (October 14--20) was 3,634, while the total during the following 7 days (October 21--27) was 4,226, an approximately 1.2-fold increase. Although this increase was smaller than the early-October surge, frequent terms during this period included academic society, NHK, and official opinion (Table~\ref{tab:event_terms}). Comparing these terms with actual news reports and expert statements showed that they corresponded to the joint statement denying shedding published by the Japanese Association for Infectious Diseases, the Japanese Respiratory Society, and the Japanese Society for Vaccinology \cite{jacso2024}, as well as media coverage by NHK and other major outlets \cite{nhk2024}. Accordingly, this study treats October 21 as the event date corresponding to the joint statement and related media coverage denying shedding.

\begin{table}[t]
\centering
\small
\caption{Top 10 frequent terms in posts around each event (7-day window, excl.\ function words and search keywords).}
\label{tab:event_terms}
\begin{tabular}{llll}
\hline\hline
\multicolumn{2}{c}{Oct.\ 1 event}
& \multicolumn{2}{c}{Oct.\ 21 event} \\
\cline{1-2}\cline{3-4}
Term & Freq.\ (\%) & Term & Freq.\ (\%) \\
\hline
harm/damage    & 0.530 & academic society & 1.272 \\
infection      & 0.526 & infection        & 1.067 \\
science        & 0.401 & NHK              & 0.808 \\
transmission   & 0.369 & official opinion & 0.637 \\
mask           & 0.327 & virus            & 0.395 \\
possibility    & 0.292 & sequelae         & 0.349 \\
denial         & 0.276 & harm/damage      & 0.335 \\
virus          & 0.260 & science          & 0.323 \\
problem        & 0.252 & countermeasure   & 0.270 \\
proof/evidence & 0.228 & recommendation   & 0.250 \\
\hline\hline
\multicolumn{4}{l}{\textit{Freq.\ = token frequency in posts within 7 days after each event.}}\\
\multicolumn{4}{l}{\textit{Function words and search keywords excluded.}}\\
\end{tabular}
\end{table}

\subsection{Repost Network and Clustering Results}

\begin{figure}[t]
  \centering
  \includegraphics[width=\columnwidth]{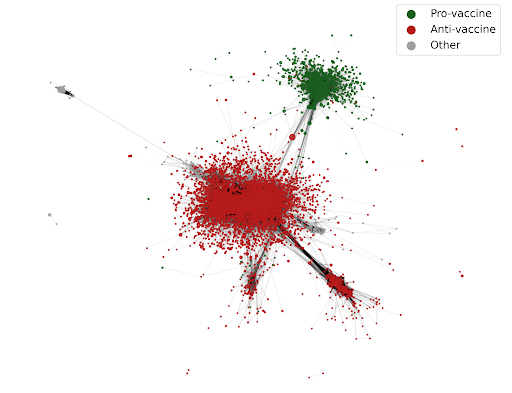}
  \caption{Visualization of the post-level repost network. Each node represents a post, and edges connect posts with similar repost-user sets based on the Simpson coefficient. Node colors indicate the interpreted discourse clusters after cluster merging: posts denying the scientific basis of shedding, posts endorsing or promoting shedding claims, and other nodes outside the two major clusters. The network exhibits a two-pole structure, with most posts concentrated in the two major clusters.}
  \label{fig:network}
\end{figure}

The resulting post-level network consisted of 15,423 nodes and 2,532,718 edges. Figure~\ref{fig:network} visualizes the network.

Louvain clustering detected 16 clusters and yielded a modularity of $Q = 0.446$, indicating a relatively clear community structure. After cluster merging based on repost-user overlap, most posts were concentrated in two major clusters, which accounted for approximately 96.6\% of all nodes. This two-pole structure is consistent with the expectation, based on prior studies, that repost behavior in vaccine-related discourse tends to be polarized by stance. The remaining 529 nodes outside the two major clusters were excluded from subsequent analyses.

One cluster contained many terms related to scientific criticism of shedding claims, such as academic society, science, conspiracy, misinformation, and evidence (Table~\ref{tab:cluster_terms}). This cluster included 2,464 posts and centered on discourse denying the scientific basis of shedding. We refer to this cluster as the pro-vaccine cluster. The other cluster contained many terms related to health damage, such as harm, sequelae, transmission, and hypersensitivity (Table~\ref{tab:cluster_terms}). This cluster included 12,430 posts and centered on discourse claiming health damage caused by shedding. We refer to this cluster as the anti-vaccine cluster.

\begin{figure}[t]
  \centering
  \includegraphics[width=\columnwidth]{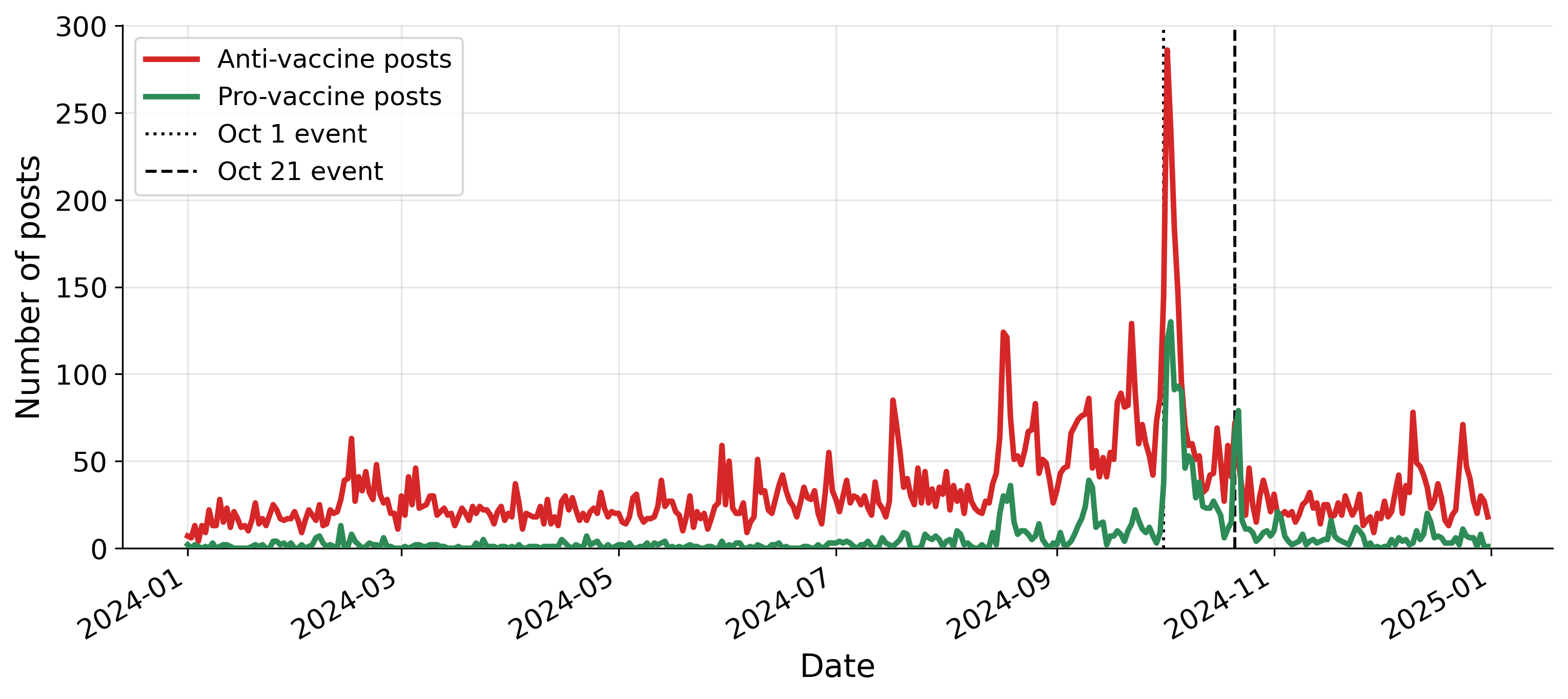}
  \caption{Daily post counts by merged repost-network cluster in 2024. The figure shows post-volume trends within the two major clusters. The anti-vaccine cluster accounted for most posts throughout the year and showed a sharp increase immediately after the October 1 event, while the pro-vaccine cluster remained relatively small but increased temporarily around the October events, particularly near the October 21 corrective response.}
  \label{fig:cluster_daily}
\end{figure}

Over the full year, the number of posts in the anti-vaccine cluster (12,430) was about five times larger than that in the pro-vaccine cluster (2,464), indicating that discourse claiming health damage from shedding dominated the shedding discourse space. Figure~\ref{fig:cluster_daily} shows the daily number of posts by cluster. The pro-vaccine cluster had low posting volume before the October events, but increased sharply after the October 21 news coverage denying shedding, temporarily exceeding the anti-vaccine cluster on October 22. The anti-vaccine cluster maintained relatively high posting volume throughout the year, but did not show a pronounced increase immediately after the denial coverage.

\begin{table}[t]
\centering
\small
\caption{Top 10 frequent terms by cluster (excl.\ function words and search keywords).}
\label{tab:cluster_terms}
\begin{tabular}{llll}
\hline\hline
\multicolumn{2}{c}{Pro-vaccine cluster}
& \multicolumn{2}{c}{Anti-vaccine cluster} \\
\cline{1-2}\cline{3-4}
Term & Freq.\ (\%) & Term & Freq.\ (\%) \\
\hline
infection        & 0.718 & harm                   & 0.882 \\
academic society & 0.490 & sequelae               & 0.439 \\
existence        & 0.465 & transmission           & 0.393 \\
mask             & 0.437 & infection              & 0.382 \\
science          & 0.407 & gas (cf.\ gaslighting) & 0.375 \\
virus            & 0.388 & psychological          & 0.337 \\
conspiracy       & 0.381 & gaslighting            & 0.328 \\
misinformation   & 0.367 & substance              & 0.311 \\
evidence         & 0.356 & medical care           & 0.307 \\
proof            & 0.293 & hypersensitivity       & 0.304 \\
\hline\hline
\multicolumn{4}{l}{\textit{Freq.\ = token frequency within cluster. Search keywords}}\\
\multicolumn{4}{l}{\textit{(vaccination, COVID-19, mRNA, etc.)\ and function words excluded.}}\\
\end{tabular}
\end{table}

\subsection{Changes in Repost Diffusion Before and After Events}
\label{sec:Changes in Repost Diffusion Before and After Events}

\begin{figure}[t]
  \centering
  \includegraphics[width=\columnwidth]{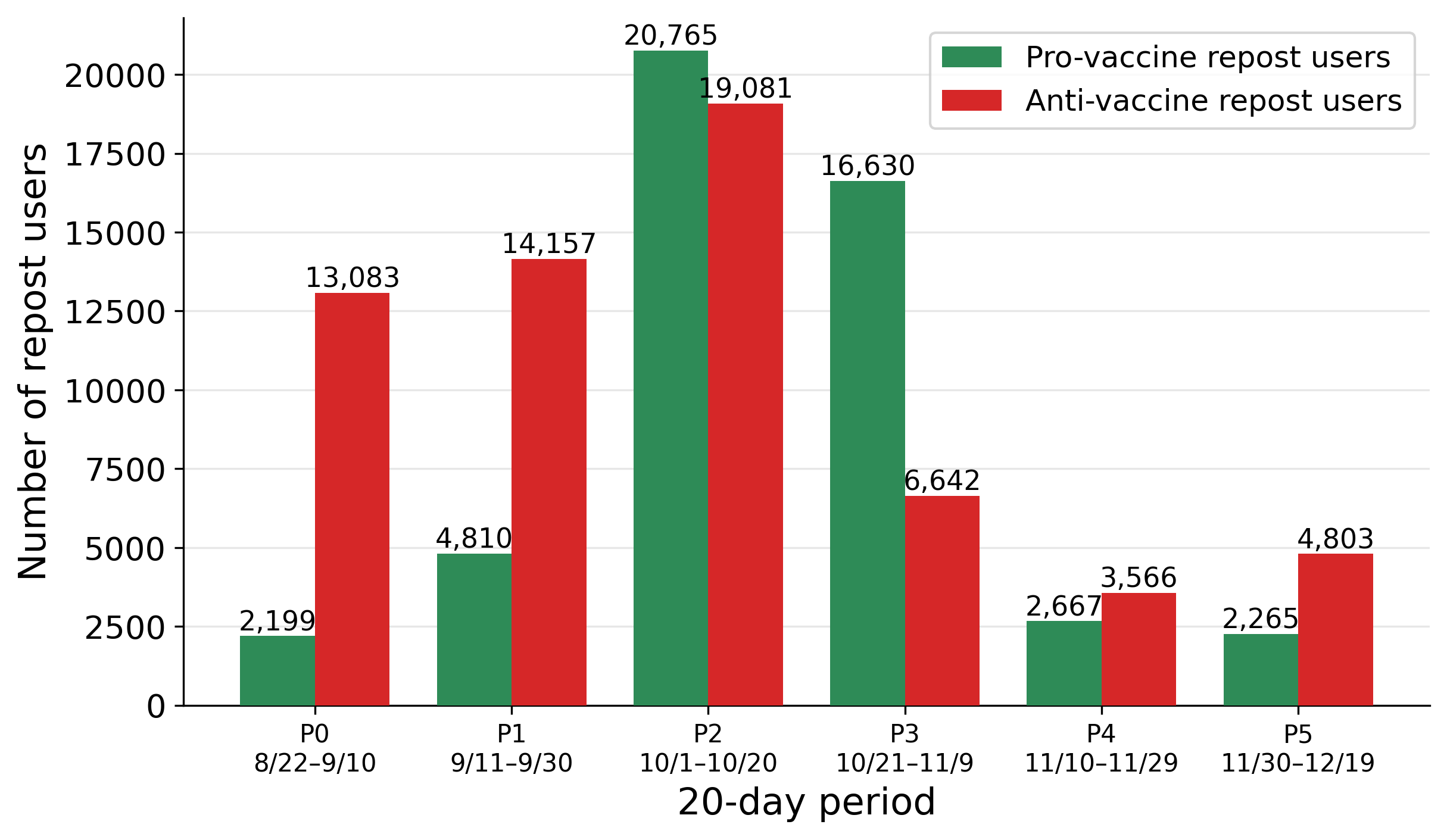}
  \caption{Number of unique repost users per 20-day period by cluster. Each bar counts unique users who reposted at least one cluster post originally published in the corresponding period. The figure illustrates the event-based comparisons reported in Tables~\ref{tab:period_users} and \ref{tab:stats}: repost participation increased sharply in P2 after the start of routine vaccination, then declined asymmetrically in P3 after the corrective response, with a sharper decrease in the anti-vaccine cluster.}
  \label{fig:period_repost}
\end{figure}

We analyzed changes in repost diffusion for the two clusters using six equal-length 20-day periods (P0--P5) anchored to the two event dates (P0: Aug.\ 22--Sep.\ 10; P1: Sep.\ 11--30; P2: Oct.\ 1--20; P3: Oct.\ 21--Nov.\ 9; P4: Nov.\ 10--29; P5: Nov.\ 30--Dec.\ 19). P2 and P3 correspond to the post-vaccination-start and post-denial-coverage phases, respectively. 
We conducted four comparisons: P1$\rightarrow$P2 and P2$\rightarrow$P3 for immediate reactions; P0+P1$\rightarrow$P3+P4 for mid-term changes; and P0+P1$\rightarrow$P4+P5 for longer-term changes.
Results are shown in Figure~\ref{fig:period_repost} and Tables~\ref{tab:period_users} and \ref{tab:stats}.

\begin{table}[t]
\centering
\small
\caption{Repost users per period by cluster.}
\label{tab:period_users}
\begin{tabular}{p{0.8cm}lrr}
\hline\hline
Period & Event context & Pro & Anti \\
\hline
P0 & Pre-event        & 2,199  & 13,083 \\
P1 & Pre-event        & 4,810  & 14,157 \\
P2 & Vaccination start      & 20,765 & 19,081 \\
P3 & Shedding denial report & 16,630 & 6,642  \\
P4 & Post-event       & 2,667  & 3,566  \\
P5 & Post-event       & 2,265  & 4,803  \\
\hline
P0+P1 & Pre-event (combined)   & 5,925  & 20,845 \\
P3+P4 & Mid-term post-event    & 17,419 & 8,495  \\
P4+P5 & Long-term post-event   & 4,130  & 6,721  \\
\hline\hline
\multicolumn{4}{p{0.9\columnwidth}}{\footnotesize Counts are non-exclusive; a user reposting both clusters within a period is counted in both columns.}\\
\end{tabular}
\end{table}

\begin{table}[t]
\centering
\small
\setlength{\tabcolsep}{4.8pt}
\caption{Statistical results for four event-based comparisons of repost diffusion.}
\label{tab:stats}
\begin{tabular}{@{}lrrrr@{}}
\hline\hline
Comparison & Pro RR & Anti RR & Anti ratio & $\chi^2(1)$ \\
\hline
P1$\to$P2       & 5.92$^{***}$  & 1.25          & 75.2\%$\to$47.7\% & 3797.3$^{***}$ \\
P2$\to$P3       & 0.531         & 0.320$^{***}$ & 47.7\%$\to$28.2\% & 2234.5$^{***}$ \\
P0+P1$\to$P3+P4 & 2.69$^{**}$   & 0.320$^{***}$ & 78.7\%$\to$32.5\% & 11136.1$^{***}$ \\
P0+P1$\to$P4+P5 & 0.665         & 0.272$^{***}$ & 78.7\%$\to$62.1\% & 1079.9$^{***}$ \\
\hline\hline
\multicolumn{5}{@{}l@{}}{
\begin{tabular}{@{}l@{}}
$^{***}$ $p < .001$, $^{**}$ $p < .01$. \\
RR = rate ratio (negative binomial regression, after vs.\ before). \\
Anti ratio and $\chi^2$ use mutually exclusive user categories.
\end{tabular}
} \\
\end{tabular}
\end{table}

In P2, after the start of routine vaccination, daily pro-vaccine repost participation increased approximately sixfold compared with P1 (RR = 5.92, $p < .001$), while anti-vaccine repost users showed only a modest and non-significant increase (RR = 1.25). As a result, the anti-vaccine ratio among mutually exclusive users decreased significantly from 75.2\,\% to 47.7\,\% ($\chi^2(1) = 3797.3$, $p < .001$).

In P3, repost participation decreased in both clusters relative to P2, reflecting the subsiding of the post-vaccination-start surge. However, the decline was asymmetric: the anti-vaccine cluster dropped more sharply (RR = 0.320, $p < .001$) while the pro-vaccine cluster showed a moderate decline that did not reach statistical significance (RR = 0.531), reducing the anti-vaccine ratio among mutually exclusive users further from 47.7\,\% to 28.2\,\% ($\chi^2(1) = 2234.5$, $p < .001$).

Comparing the pre-event period (P0+P1) with the mid-term post-coverage period (P3+P4), anti-vaccine repost participation decreased significantly (RR = 0.320, $p < .001$) while the number of pro-vaccine repost users increased significantly (RR = 2.69, $p < .01$). The anti-vaccine ratio among mutually exclusive users decreased significantly from 78.7\,\% to 32.5\,\% ($\chi^2(1) = 11136.1$, $p < .001$). This asymmetric change is difficult to explain by the overall decline of the topic alone. In the mid term, anti-vaccine diffusion participation decreased while pro-vaccine participation increased.

In the later period (P4+P5), the anti-vaccine ratio among mutually exclusive users showed signs of recovery but did not return to the pre-event level of 78.7\,\%. The anti-vaccine cluster showed a large and significant decrease relative to the pre-event period (RR = 0.272, $p < .001$), and the anti-vaccine ratio among mutually exclusive users remained significantly lower at 62.1\,\% ($\chi^2(1) = 1079.9$, $p < .001$). The pro-vaccine cluster also declined, though not significantly (RR = 0.665), indicating that the corrective information did not produce a sustained expansion of pro-vaccine diffusion.

Overall, the changes after news coverage denying shedding are best interpreted not as a one-sided sustained expansion of the pro-vaccine cluster, but as an asymmetric decline in the overall discourse activity. This decline was more pronounced in the anti-vaccine cluster, resulting in a lower share of anti-vaccine users in repost diffusion.

\subsection{Changes in User Stance Before and After Events}

We next examined which user transitions were associated with the decline in the anti-vaccine repost-user ratio after the denial response, as shown in Section~\ref{sec:Changes in Repost Diffusion Before and After Events}. We classified users' stances based on individual repost behavior and analyzed transitions before and after the events. Specifically, we analyzed mid-term transitions from P0+P1 to P3+P4. The transition analysis included 43,034 users who reposted at least one pro-vaccine or anti-vaccine cluster post in either P0+P1 or P3+P4.

\begin{figure}[t]
  \centering
  \includegraphics[width=\columnwidth]{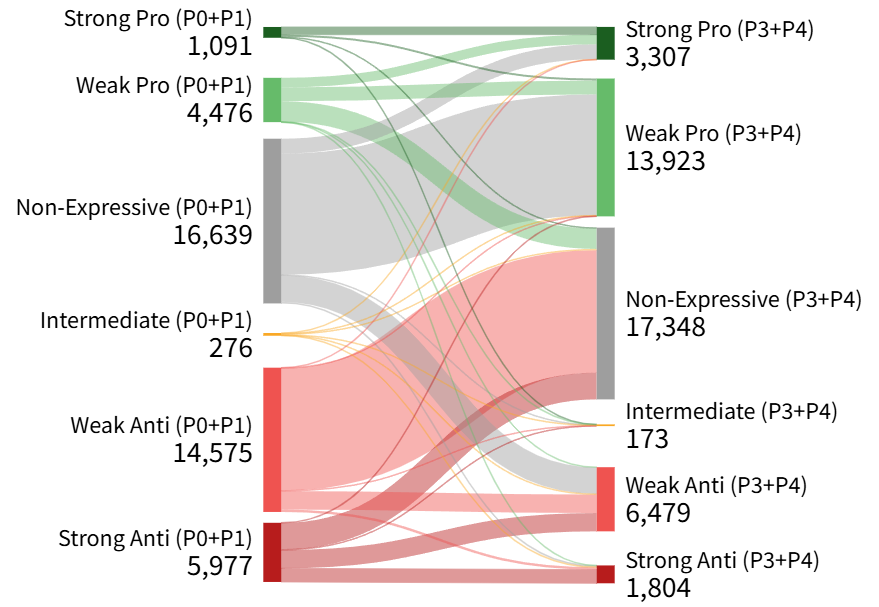}
  \caption{Sankey diagram of repost-based user stance transitions from P0+P1 to P3+P4. Nodes represent user categories defined by repost behavior, and flow thickness indicates the number of users moving between categories. The main changes were the withdrawal of anti-vaccine supporters into the non-expressive category and the entry of previously non-expressive users into pro-vaccine repost diffusion. The diagram was created using SankeyMATIC.}
  \label{fig:sankey}
\end{figure}

Figure~\ref{fig:sankey} visualizes stance transitions using a Sankey diagram. A Sankey diagram represents the amount of movement between categories by the thickness of the flows.

In the transition from P0+P1 to P3+P4, only 77 users (0.18\,\%) moved from anti-vaccine supporters to pro-vaccine supporters, while 23 users (0.05\,\%) moved in the opposite direction. Direct transitions between strong supporter groups were zero.

The number of users moving from the non-expressive group to pro-vaccine supporters was 13,742 (31.9\,\%), indicating that new participation on the pro-vaccine side persisted into the mid-term post-event period. Meanwhile, 15,071 users (35.0\,\%) moved from anti-vaccine supporters to the non-expressive group, showing substantial withdrawal from anti-vaccine diffusion participation.

\section{Discussion}

\subsection{Two-Cluster Structure of Vaccine-Related Discourse}

This study showed, through repost-network analysis, that shedding discourse formed a two-pole structure consisting of pro-vaccine and anti-vaccine clusters under a clear community structure with modularity $Q = 0.446$. This result is consistent with previous findings that repost behavior in vaccine-related discourse tends to be polarized by stance \cite{toriumi2024}. Over the full year, the number of posts in the anti-vaccine cluster (12,430) exceeded that in the pro-vaccine cluster (2,464), indicating that discourse affirming the existence of shedding dominated the shedding discourse space. 
This pattern suggests an echo-chamber-like structure in which similar groups of users repeatedly shared posts from a particular stance.

\subsection{Corrective Information Was Associated with Asymmetric Decline Rather Than Backfire}

One of the main findings of this study is that no clear diffusion-level backfire was observed in anti-vaccine repost activity after news coverage denying shedding. The results do not show that corrective information directly changed the beliefs of anti-vaccine users. Rather, after denial coverage, the overall shedding discourse declined, and repost participation decreased in both the pro-vaccine and anti-vaccine clusters. However, this decline was more pronounced in the anti-vaccine cluster, reducing the share of anti-vaccine users in overall repost diffusion.

The key point is that the decline did not occur symmetrically across the two clusters. In the short-term comparison from P2 to P3, the anti-vaccine cluster declined significantly (RR = 0.320, $p < .001$) while the pro-vaccine cluster showed a more moderate decline (RR = 0.531). Moreover, in the mid-term comparison that does not use P2 as the topic-attention peak (P0+P1$\rightarrow$P3+P4), the anti-vaccine cluster decreased significantly (RR = 0.320, $p < .001$) while the pro-vaccine cluster increased (RR = 2.69, $p < .01$). This result suggests a structural change that is difficult to explain by mere topic decline.

At the same time, the later comparison between P0+P1 and P4+P5 showed that pro-vaccine repost participation also declined, though not significantly (RR = 0.665). Therefore, the findings should not be interpreted as a sustained expansion of pro-vaccine discourse caused by corrective information. A more appropriate interpretation is that, as the overall topic declined over time, repost participation in the anti-vaccine cluster decreased more sharply, and the relative dominance of the anti-vaccine cluster remained below the pre-event level.

\subsection{Implications for Vaccine Communication on Social Media}

The transition analysis provides important implications for vaccine communication on social media. Direct transitions between strong pro-vaccine supporters and strong anti-vaccine supporters were zero, indicating that users with strong stances rarely changed their positions in response to the events. This suggests that directly persuading strong anti-vaccine core users is difficult.

At the same time, the observed movement from non-expressive users to pro-vaccine supporters (31.9\,\%) and from anti-vaccine supporters to the non-expressive group (35.0\,\%) is important. These results suggest that the main effects of corrective information may appear not as attitude conversion among strong anti-vaccine users, but as changes in information exposure and behavior among intermediate, weakly committed, and non-expressive users.

Therefore, vaccine communication on social media should not focus only on persuading strong anti-vaccine users. Instead, it is important to deliver accurate information to intermediate and non-expressive users. Corrective information may be effective less through converting core anti-vaccine users than through reaching peripheral users and suppressing new entry into anti-vaccine discourse.

The fact that the event changes appeared more clearly in repost behavior than in posting behavior also suggests that participation in information diffusion is less costly than producing original posts, making mobilization and withdrawal effects more visible in repost behavior. Evaluating the effects of public health communication should therefore consider not only post counts but also changes in diffusion participation through repost behavior.

\subsection{Limitations and Future Work}
This study has several limitations. First, it uses repost behavior as a proxy for stance, but reposts do not necessarily indicate agreement and may include critical or antagonistic resharing. Second, because the repost data did not include repost timestamps, period-based repost aggregations were assigned according to the timestamp of the original post. This may obscure the exact timing of repost diffusion. In addition, although the chi-square tests used mutually exclusive user categories within each period, the same users could appear in multiple periods. Therefore, these tests should be interpreted as descriptive comparisons of participant composition rather than as strict tests of independence. Third, the analysis is limited to posts containing the search term for ``shedding,'' and therefore does not cover the entire discourse related to shedding. Fourth, the user stance transition analysis shows aggregate changes based on period-level classifications and does not directly demonstrate individual attitude change. Finally, this study analyzes Japanese-language X discourse in 2024 and cannot determine whether the same patterns appear across other misinformation themes, platforms, or national contexts.

Future research should develop methods to distinguish supportive reposting from critical reposting by incorporating quote posts, replies, and textual context. It should also examine how different types of information sources, such as experts, media organizations, government agencies, and influencers, shape the diffusion of corrective information. Finally, similar analyses should be conducted across other vaccine misinformation themes and social contexts to examine whether the asymmetric decline observed in this study generalizes beyond the shedding discourse.

\section{Conclusion}
This study examined shedding discourse on Japanese X in 2024 through repost-network structure, diffusion changes, and user stance transitions. The discourse formed two opposing pro- and anti-vaccine clusters. Although the anti-vaccine cluster dominated overall, the joint statement and related media coverage were followed not by diffusion-level backfire but by a sharper decline in anti-vaccine repost participation.

This change was driven mainly by the withdrawal of anti-vaccine users and the entry of previously non-expressive users into pro-vaccine diffusion, rather than by the conversion of core opponents. These findings suggest that corrective information may change the composition of diffusion participants more readily than strongly held positions. Public health communication should therefore be evaluated not only by post volume but also by changes in group-specific participation.

%

\section*{Acknowledgments}
We thank Meiji Seika Pharma Co., Ltd. for its cooperation in conducting this research. Meiji Seika Pharma Co., Ltd. had no role in study design, data collection, analysis, interpretation, manuscript writing, or the decision to submit the paper. This work was supported by JST ERATO (JPMJER2502).


\end{document}